\input harvmac

\def\del{\partial }
\def \inv {^{-1}}

\def\const {{\rm const}}
\def \s {\sigma}
\def\t {\tau}
\def \ha {\half}
\def \ov {\over}

\def \four{{\textstyle {1\ov 4}}}
\def \a {\alpha}
\def\ep{\epsilon}
\def\vp {\varphi}
\def\g {\gamma}
\def \sm {$\s$-model\ }
\def \td {\tilde }

\def \lr { \lref}

\def \h {\chi}

\def \d {\del }

\def \S {{\cal S}}
\def \om {\omega}

\def\const {{\rm const}}
\def \s {\sigma}
\def\t {\tau}

\def \ha {\half}
\def \ov {\over}

\def \four{{\textstyle {1\ov 4}}}
\def \a {\alpha}
\def \lr { \lref}
\def\ep{\epsilon}

\def\vp {\varphi}
\def \bd {\bar \del}

\def\const {{\rm const}}\def\bd {\bar \del} \def\m{\mu}\def\n
{\nu}

\def\g {\gamma}

\def \sm {$\s$-model\ }
\def \td {\tilde }

\def \lr { \lref}

\def \d {\del }

\vfill\eject
\def \N {{\hat N}}

\def \h {\chi}

\def \d {\del }

\def \S {{\cal S}}

\gdef \jnl#1, #2, #3, 1#4#5#6{ { #1~}{ #2} (1#4#5#6) #3}

\lr \gibma {G.~W.~Gibbons and K.~Maeda,
``Black Holes And Membranes In Higher Dimensional
Theories With Dilaton Fields,''
Nucl.\ Phys.\ B {\bf 298}, 741 (1988).
}
\lr\silv{ A.~Adams and E.~Silverstein,
``Closed string tachyons, AdS/CFT, and large N QCD,''
hep-th/0103220.
}

\lref\horo{
G.~T.~Horowitz and K.~Maeda,
``Fate of the black string instability,''
hep-th/0105111.
}

\lref\Ur{
A.~M.~Uranga,
``Wrapped fluxbranes,''
hep-th/0108196.
}

\lref\TTU{T.~Takayanagi and T.~Uesugi,
``Orbifolds as Melvin Geometry,''
hep-th/0110099.}

\lr\gib{G.W. Gibbons,
``Quantized Flux Tubes In Einstein-Maxwell Theory And
Noncompact Internal Spaces,''
in: {\it Fields and Geometry},
ed.
A. Jadczyk (World Scientific, 1986).
G.~W.~Gibbons and K.~Maeda,
``Black Holes And Membranes In Higher Dimensional
Theories With Dilaton Fields,''
Nucl.\ Phys.\ B {\bf 298}, 741 (1988).
}

\lr \tset { A.~A.~Tseytlin,
``Harmonic superpositions of M-branes,''
Nucl.\ Phys.\ B {\bf 475}, 149 (1996)
[hep-th/9604035].
}

\lr\GS{M.~Gutperle and A.~Strominger,
``Fluxbranes in string theory,''
hep-th/0104136.
}

\lr \WWW { E.~Witten,
``Instability Of The Kaluza-Klein Vacuum,''
Nucl.\ Phys.\ B {\bf 195}, 481 (1982).
}

\lr\gal{ C.~Chen, D.~V.~Gal'tsov and S.~A.~Sharakin,
``Intersecting M-fluxbranes,''
Grav.\ Cosmol.\ {\bf 5}, 45 (1999)
[hep-th/9908132].
}

\lr \koun{ C.~Kounnas and B.~Rostand,
``Coordinate Dependent Compactifications
And Discrete Symmetries,''
Nucl.\ Phys.\ B {\bf 341}, 641 (1990).}

\lr \cone {
A.~Dabholkar,
``Strings on a cone and black hole entropy,''
Nucl.\ Phys.\ B {\bf 439}, 650 (1995)
[hep-th/9408098].
D.~A.~Lowe and A.~Strominger,
``Strings near a Rindler or black hole horizon,''
Phys.\ Rev.\ D {\bf 51}, 1793 (1995)
[hep-th/9410215].
}

\lr \suf { P.~M.~Saffin,
``Gravitating fluxbranes,''
gr-qc/0104014.
}
\lr\green{
J.~G.~Russo and A.~A.~Tseytlin,
``Green-Schwarz superstring action in a curved magnetic Ramond-Ramond
background,''
JHEP{\bf 9804}, 014 (1998)
[hep-th/9804076].
}

\lref\rohm{
R.~Rohm,
``Spontaneous Supersymmetry Breaking In Supersymmetric String
Theories,''
Nucl.\ Phys.\ B {\bf 237}, 553 (1984).
}

\lr \suy{
T.~Suyama,
``Properties of String Theory on Kaluza-Klein Melvin Background'',
hep-th/0110077.
}

\lr\oorb{
I.~R.~Klebanov and N.~A.~Nekrasov,
``Gravity duals of fractional branes and logarithmic RG flow,''
Nucl.\ Phys.\ B {\bf 574}, 263 (2000)
[hep-th/9911096].
M.~Bertolini, P.~Di Vecchia, M.~Frau, A.~Lerda, R.~Marotta and
I.~Pesando,
``Fractional D-branes and their gauge duals,''
JHEP {\bf 0102}, 014 (2001)
[hep-th/0011077].
J.~Polchinski,
``N = 2 gauge-gravity duals,''
Int.\ J.\ Mod.\ Phys.\ A {\bf 16}, 707 (2001)
[hep-th/0011193].
}

\lr\magnetic{
J.~G.~Russo and A.~A.~Tseytlin,
``Magnetic flux tube models in superstring theory,''
Nucl.\ Phys.\ B {\bf 461}, 131 (1996)
[hep-th/9508068].}

\lr\heterotic{
J.~G.~Russo and A.~A.~Tseytlin,
``Heterotic strings in uniform magnetic field,''
Nucl.\ Phys.\ B {\bf 454}, 164 (1995)
[hep-th/9506071].}

\lr\ourold{
J.~G.~Russo and A.~A.~Tseytlin,
``Exactly solvable string models of curved space-time backgrounds,''
Nucl.\ Phys.\ B {\bf 449}, 91 (1995)
[hep-th/9502038].}

\lref\duff{
M.~J.~Duff and J.~X.~Lu,
``The Selfdual type IIB superthreebrane,''
Phys.\ Lett.\ B {\bf 273}, 409 (1991).
}

\lr \ktmn{
I.~R.~Klebanov and A.~A.~Tseytlin,
``Gravity duals of supersymmetric SU(N) x SU(N+M) gauge theories,''
Nucl.\ Phys.\ B {\bf 578}, 123 (2000)
[hep-th/0002159].
J.~M.~Maldacena and C.~Nunez,
``Towards the large N limit of pure N = 1 super Yang Mills,''
Phys.\ Rev.\ Lett.\  {\bf 86}, 588 (2001)
[hep-th/0008001].
}

\lref\ada{
A.~Adams, J.~Polchinski and E.~Silverstein,
``Don't panic! Closed string tachyons in ALE space-times,''
hep-th/0108075.
}

\lr\gau{ F.~Dowker, J.~P.~Gauntlett, D.~A.~Kastor and
J.~Traschen,
``Pair creation of dilaton black holes,''
Phys.\ Rev.\ D {\bf 49}, 2909 (1994)
[hep-th/9309075].
F.~Dowker, J.~P.~Gauntlett, S.~B.~Giddings and G.~T.~Horowitz,
``On pair creation of extremal black holes and Kaluza-Klein
monopoles,''
Phys.\ Rev.\ D {\bf 50}, 2662 (1994)
[hep-th/9312172].
}

\lr\gauu{
F.~Dowker, J.~P.~Gauntlett, G.~W.~Gibbons and
G.~T.~Horowitz,
``The Decay of magnetic fields in Kaluza-Klein theory,''
Phys.\ Rev.\ D {\bf 52}, 6929 (1995)
[hep-th/9507143].
}

\lr\DG{ F.~Dowker, J.~P.~Gauntlett, G.~W.~Gibbons and
G.~T.~Horowitz,
``Nucleation of $P$-Branes and Fundamental Strings,''
Phys.\ Rev.\ D {\bf 53}, 7115 (1996) [hep-th/9512154].
}

\lr\GLF{R.~Gregory and R.~Laflamme,
``Black strings and p-branes are unstable,'' Phys.\ Rev.\ Lett.
\ {\bf 70}, 2837 (1993) [hep-th/9301052]. }

\lr\closed{A.~A.~Tseytlin,
``Closed superstrings in magnetic field:
instabilities and supersymmetry breaking",
Nucl.\ Phys.\ Proc.\ Suppl.\ {\bf 49}, 338 (1996)
[hep-th/9510041].
}
\lr\tse{A.~A.~Tseytlin,
``Melvin solution in string theory,''
Phys.\ Lett.\ B {\bf 346}, 55 (1995)
[hep-th/9411198].
}

\lr \CG {M.~S.~Costa and M.~Gutperle,
``The Kaluza-Klein Melvin solution in M-theory,''
JHEP {\bf 0103}, 027 (2001)
[hep-th/0012072].
}

\lr\ues{
T. Takayanagi and T. Uesugi,
``Orbifolds as Melvin Geometry'',
hep-th/0110099.}

\lref\Emp{
R.~Emparan,
``Composite black holes in external fields,''
Nucl.\ Phys.\ B {\bf 490}, 365 (1997)
[hep-th/9610170].
}

\lr \gan{ A.~Bergman, K.~Dasgupta, O.~J.~Ganor, J.~L.~Karczmarek
and G.~Rajesh, ``Nonlocal field theories and their gravity
duals,'' hep-th/0103090.
}

\lref\matm{
Y.~Sekino and T.~Yoneya,
``From supermembrane to matrix string,''
hep-th/0108176.
L.~Motl,
``Melvin matrix models,''
hep-th/0107002.
}

\lr\chep{
I.~Chepelev and A.~A.~Tseytlin,
``Long-distance interactions of branes: Correspondence between
supergravity and super Yang-Mills descriptions,''
Nucl.\ Phys.\ B {\bf 515}, 73 (1998)
[hep-th/9709087].
}

\lr\RTN{ J.~G.~Russo and A.~A.~Tseytlin,
``Magnetic backgrounds and tachyonic instabilities in closed
superstring theory
and M-theory,''
hep-th/0104238.
}
\lr\RTT{
A.~A.~Tseytlin,
``Magnetic backgrounds and tachyonic instabilities in closed string
theory,''
hep-th/0108140.
}

\lr\fip{
J.~Figueroa-O'Farrill and G.~Papadopoulos,
``Homogeneous fluxes, branes and a maximally supersymmetric solution of
M-theory,''
JHEP {\bf 0108}, 036 (2001)
[hep-th/0105308].
}

\lr \duffp{M.~J.~Duff, H.~Lu and C.~N.~Pope,
``Supersymmetry without supersymmetry,''
Phys.\ Lett.\ B {\bf 409}, 136 (1997)
[hep-th/9704186].
}

\lr \KR {
C.~Kounnas and B.~Rostand,
``Coordinate Dependent Compactifications And Discrete Symmetries,''
Nucl.\ Phys.\ B {\bf 341}, 641 (1990).
}


\baselineskip8pt
\Title{\vbox
{\baselineskip 6pt{\hbox{}}{\hbox
{}}{\hbox{hep-th/0110107}} {\hbox{
}}} }
{\vbox{\centerline {
Supersymmetric fluxbrane intersections }
\vskip4pt
\centerline{and closed string tachyons }
}}

\vskip -30 true pt

\centerline { {J.G. Russo$^a$ \footnote {$^*$} {e-mail address:
russo@df.uba.ar } and A.A. Tseytlin$^b$
\footnote{$^{\star}$}{\baselineskip8pt e-mail address:
tseytlin.1@osu.edu } \footnote{$^{\dagger}$}{\baselineskip8pt Also
at Imperial College, London and Lebedev Physics Institute,
Moscow.} }}

\smallskip \smallskip

\centerline {\it {}$^a$ Departamento de F\'\i sica, Universidad de
Buenos Aires, }
\smallskip
\centerline {\it Ciudad Universitaria, Pab. I, 1428 Buenos Aires,
Argentina
}

\medskip

\smallskip\smallskip
\centerline {\it {}$^b$ Smith Laboratory, Ohio State University, }
\smallskip

\centerline {\it Columbus OH 43210-1106,
USA }
\bigskip

\centerline {\bf Abstract}
\medskip
\baselineskip10pt
\noindent

\noindent
We consider NS-NS superstring model
with several ``magnetic'' parameters
$b_s$ ($s=1, ...,N $) associated with twists mixing
a compact $S^1$ direction
with angles in $N$ spatial 2-planes
of flat 10-dimensional space. It generalizes the Kaluza-Klein
Melvin model which has  single parameter $b$.
The corresponding
 U-dual background is  a R-R type IIA solution
describing an orthogonal intersection of $N$ flux 7-branes.
Like the Melvin model, the
NS-NS string model with $N$ continuous parameters
is explicitly solvable; we present its perturbative
spectrum and torus partition function explicitly
for the $N=2$ case.
For generic $b_s$ (above some critical values)
there are tachyons in the $S^1$ winding
sector.  A remarkable feature of this model
is that while in the Melvin $N=1$ case all
supersymmetry is broken,
a fraction of it may be preserved for $N >1$ by making a
special choice of the parameters $b_s$.
Such solvable NS-NS models may
be viewed as continuous-parameter
analogs of non-compact orbifold
models; they and their U-dual R-R fluxbrane counterparts
may have some ``phenomenological'' applications.
In particular, in $N=3$ case one finds a special
1/4 supersymmetric R-R 3-brane background.
Putting Dp-branes in flat twisted  NS-NS  backgrounds
leads to world-volume gauge theories with reduced
amount of supersymmetry.
We also discuss possible evolution patterns
of unstable backgrounds
towards stable supersymmetric ones.

\medskip

\Date {October 2001}

\noblackbox
\baselineskip 15pt plus 2pt minus 2pt

\lr\tsem{
A.~A.~Tseytlin,
``Melvin solution in string theory,''
Phys.\ Lett.\ B {\bf 346}, 55 (1995)
[hep-th/9411198].
}

\lr\ourold{
J.~G.~Russo and A.~A.~Tseytlin,
``Exactly solvable string models of curved space-time backgrounds,''
Nucl.\ Phys.\ B {\bf 449}, 91 (1995)
[hep-th/9502038].
}

\lref\Chen{
C.~Chen, D.~V.~Gal'tsov and S.~A.~Sharakin,
``Intersecting M-fluxbranes,''
Grav.\ Cosmol.\ {\bf 5}, 45 (1999)
[hep-th/9908132].
}

\lref\Costa{
M.~S.~Costa, C.~A.~Herdeiro and L.~Cornalba,
``Flux-branes and the dielectric effect in string theory,''
hep-th/0105023.
R.~Emparan,
``Tubular branes in fluxbranes,''
Nucl.\ Phys.\ B {\bf 610}, 169 (2001)
[hep-th/0105062].
D.~Brecher and P.~M.~Saffin,
``A note on the supergravity description of dielectric branes,''
Nucl.\ Phys.\ B {\bf 613}, 218 (2001)
[hep-th/0106206].
}

\newsec{Introduction}

Recently, there  has been  a renewed interest in studying
magnetic backgrounds analogous to Melvin fluxtubes
(see \refs{ \gib\gau\tsem\ourold\gauu\magnetic\DG
\Emp
\green\Chen\CG\suf\GS\RTN\Costa\Ur\RTT\matm\suy-\TTU }).
Such backgrounds continuously interpolate between tachyonic and
tachyon-free non-supersymmetric string vacua.
They are also smoothly connected to the
supersymmetric closed string vacua.
They provide a simple framework to study the decay of unstable
backgrounds by classical and quantum (non-perturbative) effects.
In particular, one can compute the semiclassical amplitude for a decay
of spacetime by instanton effects \refs{\gauu, \CG }\ into stable
supersymmetric configurations, or
study the instability of type 0A string
within the effective field theory and supergravity \RTN .


The Kaluza-Klein Melvin magnetic fluxtube background
\refs{\gib,\gauu,\DG}
and the corresponding string theory \refs{\ourold,\magnetic}
are, in general,
non-supersymmetric
(except for the ``trivial'' magnetic field values $b_s$,
when the string
theory reduces to the standard
flat superstring theory).
Remarkably,
the direct generalization of
the KK Melvin background to the case with
several magnetic
parameters $b_s$ ($s=1,..., N\leq 4$),
considered previously at the
supergravity level in \refs{\DG, \Ur},
preserves a fraction of supersymmetry for
special choices of $b_s$
(this generalizes the observation made in \GS\
in the case of $N=2$).
As we shall demonstrate
below,
this produces a supersymmetric model
which has $N-1$ {\it continuous} parameters.

This NS-NS string model
combines the simplicity (solvability)
with supersymmetry without imposing a
relation between the radius $R$ of the KK direction and the
magnetic $b_s$ parameters.
For example, the
$N=2$ model is tachyon-free for any radius only in
the supersymmetric limit $b_1= b_2$, while
the 3-parameter model
is supersymmetric for $b_1= b_2+b_3$
(modulo trivial sign changes).

The supersymmetry of special
multi-parameter Melvin-type models
is related to some previously known facts.
In the case of the rational
choice of the twist parameters $b_s R= 1/n_s$
($R$ is the KK radius)
these models look similar \refs{\GS,\Ur}
(but are not, in fact, equivalent)
to the standard
supersymmetric C$^{N}/Z_n$
orbifold models (with the $N=2$ case discussed
recently in \ada).
For generic $b_sR$, the supersymmetric
$b_1=b_2$ case of the $N=2$ NS-NS background is
U-dual of the
supersymmetric R-R type IIA ``F5-brane''
background of \GS\ (see also \Ur).


This paper is organized as follows.
In section 2 we present the NS-NS superstring model, and
describe the cases where a fraction of supersymmetry is preserved.
In section 3 we solve the corresponding
conformal string model in
terms of free fields, determine the string mass spectrum
and compute the one-loop partition function.
As expected, the latter
vanishes in the supersymmetric cases, is finite in the
non-tachyonic non-supersymmetric ones,
is trivial
and is IR divergent in the cases where tachyons appear
in the spectrum.
We discuss the appearance of tachyons, and possible outcomes of
evolution of the unstable backgrounds.
We also consider
a more general curved-space background
with parameters $b_s, \tilde b_s$,
which contains as special cases the original flat KK
background and its T-dual counterpart.
The corresponding
non-trivial NS-NS string model is an exact
conformal field theory to all orders in $\a'$,
and is again solvable in terms of free fields.

In section 4 we describe U-dual
supersymmetric R-R fluxbrane
backgrounds, where the magnetic field (with fluxes in
$N$ different planes)
corresponds to the R-R one-form potential
of type IIA supergravity.
These ``F$p$-branes'', with $p=9-2N$, have proper interpretation as
orthogonal intersections \DG\ of $N$ F7-branes.
The magnetic field
parameters $b_s$ control the strength of supersymmetry breaking,
with the supersymmetric limit being $b_1= b_2+...+b_{N} $, for
$N=2,3,4$.
In particular, one finds, in addition to the
``F5-brane'' with 16 supersymmetries \GS,
an ``F3-brane'' with 8 supersymmetries, and an ``F1-brane''
with 4 supersymmetries.
We argue that non-supersymmetric fluxbrane
backgrounds should decay
quantum mechanically via non-perturbative instanton effects
into the stable supersymmetric fluxbranes.

In section 5 we discuss  some
possible  applications, and,
 in particular,  the structure of  the  $\cal N$=2, $d=4$
  world-volume
 theory corresponding to the F3-brane.
In Appendix  we present   supergravity solutions
which generalize the standard  Dp-branes
to the case  when
the flat transverse space  contains
magnetic twists which reduce the amount of
world-volume supersymmetry.

\newsec{The NS-NS superstring model}

\subsec{Definition of the model and
conditions of supersymmetry}

The string model we are going to consider is a direct
generalization of the KK Melvin model \magnetic , where
the corresponding background \refs{\gau, \gauu }
was a flat space with one compact coordinate $x_9$ ``mixed''
with the polar angle of a spatial 2-plane.
Namely, let us select
$N$ spatial 2-planes and combine the rotations
in them with shifts around the KK circle \DG.
The resulting metric is
\eqn\melpp{
ds^2_{10}=-dt^2+dx_i^2+dy^2+\sum_{s=1}^N \big[dr_s^2+ r_s^2 (d\varphi_s
+ b_s dy)
^2\big]
\ ,
}
where
$$
y\equiv x_9 \ ,\ \ \ \ \ \ \
i=2N+1,...,8\ ,\ \ \ \ N=1,2,3,4\ ,
$$
$$
x_9\equiv x_9+2\pi R\ , \ \ \ \ \ \ \ \vp_s\equiv \vp_s+2\pi\ .
$$
Though locally flat, this metric is topologically trivial only
if
\eqn\kop{
b_s R =m_s\ , \ \ \ \ \ \ \ \ \ \
m_s = 0, \pm 1, \pm 2, ... \ ,\ \ \ s=1,...,N\ .
}
A similar construction can be carried out in eleven dimensions
leading, upon dimensional reduction,
to curved 10-d space-time geometries
with R-R magnetic flux (see \refs{\DG,\green,\GS,\Ur} and section 4).

We will be interested, in particular,  in the NS-NS string model
defined by
the locally flat 10-d metric \melpp\  with
 $N=2$
\eqn\melv{
ds^2_{10}=-dt^2+dx_i^2+dy^2+dr_1^2+ r_1^2 (d\varphi_1 + b_1 dy) ^2
+ dr_2^2 + r_2^2 (d\varphi_2 + b_2 dy)^2 \ .
}
Since going around the KK circle $x_9$ is accompanied by
the rotation in the two planes \
exp$ [2\pi i( b_1 R J_{12} + b_2 R J_{34})]$,
the supersymmetry is trivially preserved
if the integers $m_1,m_2$ in \kop\ are even, i.e. if
$b_sR=2n_s$.
In this case, the string model
is equivalent to the standard superstring in flat space
(the same is obviously true for any $N$).

Let us now look for the cases when
part of supersymmetry may be
preserved. Since the space is
flat (and we assume that all other supergravity fields
have trivial backgrounds)
the standard Killing spinor condition
\eqn\kili{
(\del_\mu + \four
\om^{mn}_{ \m }\g_{mn}) \ep =0 \
}
is equivalent to the condition of existence of
residual global symmetry
in the Green-Schwarz string action.
We shall follow closely the discussion in \magnetic.
The
light-cone gauge GS Lagrangian
takes a very simple form when the background geometry is flat
\eqn\gss{ L_{GS}
= L_B + L_F= G_{\m\n} (x)\del_+ x^\m \del_- x^\n
+ i \S_R {\cal D}_+ \S_R
+ i \S_L {\cal D}_- \S_L\ , } $$ {\cal D}_a\equiv \del_a + \four
\om^{mn}_{ \m }\g_{mn}\del_a x^\m
\ . $$
Here $\S^p_{R,L}$ $\ (p=1,...,8)$ are the right and left real
spinors of
$SO(8)$ (we consider type IIA theory).

To keep the discussion general, let us assume that
the spatial
part of the bosonic term in the action
has the form ($y=x_9$)
\eqn\mes{
L_B = (\d_+ x_m - f_{mn} x_n \d_+ y) (\d_- x_m - f_{mk} x_k \d_- y) +
\d_+ y \d_- y \ , }
where $m,n,k=1,2,..., 2N$ and
$f_{mn}$ is a {\it constant }
antisymmetric matrix.
This may be interpreted as an
action for a set of 2-d scalar fields coupled to a
locally-trivial 2-d
gauge potential $(A_a)_{mn}= f_{mn} \del_a y$
taking values in the algebra of the
$SO(2N)$ rotation group.
In the simplest $N=1$ case
we have $f_{mn}= b \ep_{mn}, \ m,n=1,2$.
In general,
we may choose the coordinates so that
$f_{mn}$ takes a block-diagonal form:
$$f_{s, s+1}=-f_{s+1,s} = b_s\ ,\ \ \ \ \ \ \ \ \ \
s=1,2,...,N\ . $$
$N$ (which may be 1,2,3 or 4)
is the number of 2-planes where $f_{mn}$ has non-zero components.

In the natural vierbein basis ($e_m = dx_m - f_{mn} x_n dy, \
\ e_y=
dy$) the spin
connection 1-form is
$\om^{mn} =- f_{mn}dy, \ \om^{m,y} =0$
so that the fermionic part of
\gss\ becomes
\eqn\ongs{L_{F} = i \S_R (\del_+ - \four f_{mn} \g_{mn}\d_+ y )
\S_R
+ i \S_L (\del_- - \four f_{mn} \g_{mn} \d_- y ) \S_L
\ . }
To look for a residual global supersymmetry $ \S \to \S + \ep$
we are thus to solve
the zero-mode (Killing-spinor) equation \kili\ for a space-time
spinor $\ep= \ep (x^i,y)$:
\eqn\kiii{ (\del_y - \four f_{mn}\g_{mn} ) \ep =0 \ . }
The formal solution is
\eqn\kii { \ep (y) = \exp (\four f_{mn}\g_{mn} y)\ \ep_0 \ , \ \ \
\ \ \ \ep_0 = \const \ . }
It does not, in general, satisfy the necessary
periodic boundary condition in $y$,
$\ \ep (y + 2\pi R)= \ep (y)$,
unless $f_{mn}$ is such that
\eqn\fef{ e^{ { 1 \ov 2} \pi R f_{mn}\g_{mn}} \ep_0 = \ep_0 \ }
has non-trivial solutions for $\ep_0$.
In the simplest $N=1$
case, when $f_{mn}$ has just one non-zero eigen-value
$f_{12}=b$,
the condition \fef\ becomes
$ \exp (\pi b R \g_{12}) \ep_0 =
[ \cos (\pi b R) + \g_{12} \sin (\pi b R) ] \ep_0 ,
$
so that the supersymmetry is preserved only if $bR= 2n $.
In this case the model becomes trivial, i.e.
equivalent to the flat space superstring.

In the case of $N \geq 2$ non-zero eigenvalues
of $f_{mn}$ the
full supersymmetry is again trivially
preserved if $ b_s R = 2 n_s$
($s=1, ..., N$).\foot{
The parameters $b_s$ are, in fact, defined modulo shifts by
$2 n_s/R$, so non-trivial string models can be parametrized, e.g.,
by
$ 0 < b_s < 2/R$.
We shall assume this in what follows.}
But now there is also a possibility of preserving a fraction
of supersymmetry without relating $R$ to $b_s$ but instead
imposing a condition on $b_s$ and
choosing a special solution for $\ep_0$.
Indeed, let us look for solutions of
\eqn\proj{ f_{mn}\g_{mn} \ep_0 = 0 \ . }
This problem is isomorphic to the condition of preservation of a
fraction of supersymmetry in (abelian) gauge field theory in $2N$
dimensions
in a magnetic gauge field background,
i.e. to the condition of vanishing of the gaugino variation.
In the case of two non-zero eigenvalues $b_1,b_2$
one finds that 1/2 of supersymmetry is preserved if the configuration
is
selfdual $b_1= b_2$ (or anti-selfdual $b_1=- b_2$).
Indeed, in this case we get from \proj\
\eqn\proo{(\g_{12} + \g_{34}) \ep_0 = 0 \ , }
which is a projector relation
\eqn\pou{
P_- \ep_0 = 0 \ , \ \ \
P_+ \ep_0 = \ep_0 \ , \ \ \ \ \ \ \ \ \ \
P_\pm = \ha ( 1 \pm \gamma_{1234}) \ , \ \ \ P_+ + P_- =1 \ . }
More generally, if we assume that $\ep_0$ solves \proo\
then the periodicity condition \fef\ becomes
\eqn\kiir{ e^{ \pi R (b_1 - b_2) \g_{12} }\ep_0 = \ep_0
\ ,
}
so that supersymmetry -- a $y$-periodic Killing spinor --
exists if
\eqn\kok{
R(b_1 - b_2) = 2n = 0, \pm 2, \pm 4, ... \ . }
Under the assumption
$ 0 < Rb_s < 2$ that leaves us with
$
R(b_1 -b_2) = 0$, i.e. $b_1 =b_2$ as the only {\it non-trivial} option.
In this case 1/2 of 32 original supersymmetries is preserved.

For three non-zero eigen-values
(twists in three planes) we get from \proj\
the following condition (multiplying by $\g_{12}$)
\eqn\nono{
( 1 - { b_2 \ov b_1} \g_{1234} - { b_3 \ov b_1}
\g_{1256}) \ep_0 = 0 \ .
}
While the operator here is never a projector,
this equation can be satisfied if we set, e.g.,
\eqn\sii{
b_1 = b_2 + b_3\ ,
}
and impose the following two conditions
(two copies of the ``self-duality'' conditions that
are realized by projectors)
\eqn\koo{
(\g_{12} + \g_{34}) \ep_0 = 0 \ , \ \ \ \ \
(\g_{12} + \g_{56}) \ep_0 = 0 \ .
}
Other possibilities are obviously equivalent to this one
by changing signs and
renaming the parameters.
The resulting solution preserves 1/4 of supersymmetry.\foot{Related
gauge field configuration
appears in SYM theory on a 6-torus (see e.g. \chep).}
More general solutions are obtained
by imposing \koo\ and then solving the
periodicity condition \fef:
$e^{ \pi R (b_1 - b_2 - b_3 ) \g_{12} } \ \ep_0
=\ep_0$,
i.e. \
\eqn\kokk{
R (b_1 - b_2 - b_3 ) = 2n = 0, \pm 2, \pm 4, ... \ . }
By virtue of the periodicity in $b_s$ ($b_s \equiv b_s + 2n_s R^{-1}$),
the resulting theory is equivalent
to the one with $b_s$ related by \sii .

Let us note that for $b_s$ in the ``fundamental domain''
$ 0< Rb_s < 2 $,
the condition \kokk\ is formally solved not only by \sii, but also by
\eqn\peerr{
R (b_1 - b_2 - b_3 ) = -2 \ .
}
While the resulting string theory is equivalent, as mentioned
above,
to the one defined by \sii, here
the Killing spinor has a non-trivial dependence
on $y$, i.e. is not constant along the KK direction.
This
implies that upon dimensional reduction in $y$
or T-duality in $y$ the
supersymmetry will not be preserved
at the {\it supergravity} level
(but will of course be preserved at the level of the
perturbative {\it string} theory).\foot{In the case of
reduction from 11 to 10 dimensions one again will have
loss of supersymmetry at the level of supergravity
(and, in this case, also at the level of
perturbative string theory in the
corresponding R-R background).
However, supersymmetry which is present in the full
11-d (membrane) theory should be restored
in the 10-d string theory once non-perturbative
string states are taken into account
(see \duffp\ for a related discussion).}

We conclude that in the case of the two and more twist parameters
it is possible to preserve a fraction
of supersymmetry without imposing a relation between
the KK radius $R$ and the twists.
The resulting non-compact NS-NS string models
with continuous parameters
provide a simple laboratory for study of
issues of supersymmetry breaking
and tachyons in closed string theory.

In the special case of rational parameters, e.g.,
$b_s R = 1/n_s$, these models look similar \refs{\tsem,\GS,\Ur}
to the non-compact
orbifold models or strings on the cone \refs{\cone}.
Indeed, in the $N=1$ case
the coordinate $\vp'= \vp + b y$ in \melpp\
becomes $2\pi/n$ periodic. Introducing
the new $2\pi$-periodic coordinates
$\td \vp= n \vp' $ and the new $2\pi R$ periodic coordinate
$\td y = R \vp$
one finds that the metric becomes
\eqn\elcon{
ds^2_{10}=-dt^2+dx_i^2+ n^2 (d\td y - { 1\ov n} d \td \vp)^2
+ dr^2+ { r^2\ov n^2} d\td \vp^2
\ . }
The 2-plane part of this metric is indeed
a metric of a cone, but in addition there is a non-vanishing
KK vector. Thus the metric cannot be represented as a direct
product of a cone and a circle, and its non-trivial 3-d part is,
in fact, {\it non}-singular
(the metric \melpp\ has no conical singularities
for any $b_s R$).
As a result, the corresponding string model
(in particular, its spectrum)
is not equivalent to
the one for the
direct product of an orbifold $C/Z_n $ and a circle $ S^1$.
There is, however, a close similarity, in particular regarding
instabilities and the presence of supersymmetry for $N > 1$
for a special choice of $n_s$.
For example, in the $N=2$ case
the C$^2/(Z_{n_1} \times Z_{n_2} )$ orbifold is also supersymmetric if
$n_1 = \pm n_2$, i.e. $b_1 =\pm b_2$ \refs{\ada,\GS}.

\newsec{
Solution of the
$(b_1,b_2)$ NS-NS string model}
\subsec{Free field representation }

It is possible to express the bosonic $x^m$
and fermionic
${\cal S}_{L,R}$ coordinates in \mes,\ongs\ terms of free fields.
For simplicity, we shall explicitly
consider the case of the
two non-zero eigenvalues
$b_1=f_{12}$,
$b_2=f_{34}$.
The Lorentz group $SO(8)$ is then broken to:
$$
SO(8)\to SO(4)\times SO(2)\times SO(2)\ ,
$$
where the two factors $SO(2)$ represent rotations in the 1-2 and
3-4 plane, respectively.
Fermion representations decompose as follows:
$$
{\cal S}_R=\psi_R^{++}\oplus \bar \psi_R^{--}\oplus\psi_R^{+-}\oplus
\bar \psi_R^{-+}\ ,
$$
$$
8_R\to (2_R,\ha ,\ha )\oplus(\bar 2_R,-\ha ,-\ha )\oplus(2_R,\ha ,-\ha
)
\oplus(\bar 2_R,-\ha
,\ha )
$$
and the same for $ {\cal S}_L$. The bosonic and fermionic parts of
the GS lagrangian \mes,\ongs\ take the form:
$$
L_{B}=\del_+ x_i \del_- x_i + (\del_+ +ib_1 \del_+ y)z_1\ (\del_-
-ib_1
\del_- y)z_1^*
\ ,
$$\eqn\geer{
+\ (\del_+ +ib_2 \del_+ y)z_2\ (\del_- -ib_2 \del_- y)z_2^* + \del_+ y
\del_- y
\ , }
$$z_1\equiv x_1+ix_2 \ , \ \ \ \ \ z_2\equiv x_3+ix_4 \ , $$
and
$$
L_{F}= i\bar \psi_R^{--} [\del_+ + \ha i ( b_1 +b_2) \del_+ y
]\psi_R^{++}
+i\bar \psi_R^{-+} [\del_+ + \ha i ( b_1 -b_2) \del_+ y ]\psi_R^{+-}
$$
\eqn\eer{
+\ i\bar \psi_L^{--} [\del_- + \ha i ( b_1 +b_2) \del_- y ]\psi_L^{++}
+
\ i\bar \psi_L^{-+} [\del_- + \ha i ( b_1 -b_2) \del_- y ]\psi_L^{+-}
\ . }
The Lagrangian can be written in terms of redefined bosons and
fermions which are free fields
$$
z_1= e^{i b_1 y}Z_1\ ,\ \ \ \ \ \ \ \ z_2=e^{i b_2 y} Z_2\ ,\
$$
$$
Z_1(\s+\pi)=e^{2i \pi b_1 wR} Z_1(\s )\ ,\ \ \ \ \ \ \ Z_2
(\s+\pi)=e^{2i \pi
b_2 wR} Z_2(\s )
\ ,
$$
\eqn\free{
\psi_R^{++}= e^{ - {i\over 2} (b_1+b_2) y} \psi_{0R}^{++} \ ,\ \ \ \
\psi_{0R}^{++}(\sigma +\pi )= e^{i \pi w R (b_1+b_2)}
\psi_{0R}^{++}(\sigma )
\ , \ \ \ etc., }
where $w$ is the winding number in the $y$ direction.

{}{}From these equations it is easy to see
why the case of $b_1=b_2$ (or $b_1=-b_2$) is supersymmetric.
Then the fermions $\psi ^{+-}_{R,L}$ and $\bar \psi^{-+}_{R,L}$
are decoupled from $y$,
i.e. are free fields that do not transform under $y\to y+2\pi R$.
There are four free Weyl fermions
with nontrivial periodicity:
$\psi^{++}_{0L},\ \psi^{++}_{0R},\ \bar \psi^{--}_{0L},\ \bar
\psi^{--}_{0R}$.
But they have the same boundary
conditions as the bosonic degrees of freedom
$Z_{1L},Z_{1R}, Z_{2L}, Z_{2R}$.
Thus, for every fermionic degree of freedom,
there is a bosonic degree of freedom with the same periodicity.
This 2-d supersymmetry is correlated with the space-time one
in the light-cone gauge GS description.

An interesting special
case is when $b_1=b_2=R^{-1} (n+\ha) $.
Here there are
free fermions obeying antiperiodic boundary conditions in $y=x_9$
in the odd winding number sector, but
at the same time the supersymmetry is preserved since
there are also antiperiodic
free bosons $Z_1,Z_2$.\foot{
The physical fermion and boson coordinates
in \geer\ and \eer\ do not of course transform
under shifts of $x_9 $.}
Note that for a generic value of $b_1=b_2$
this model is not equivalent to the standard
(periodic or antiperiodic) free superstring theory
in flat space. The non-triviality of
the model expressed in terms of the free fields
is caused by the special boundary conditions.\foot{This
class of models
is closely related to the
to the Scherk-Schwarz type compactifications
in string theory \refs{\rohm,\KR}.}

\subsec{Mass spectrum }

This superstring model can be solved by a simple
generalization
of the discussion in the case of $b_1\not=0, b_2=0$ in \magnetic .
Let $\hat N_R$ and $\hat N_L$ denote the number of states operators,
which
have the same form as in the free superstring theory.
They have integer non-negative eigenvalues in the GS description,
while
in the NSR approach they
are expressed in terms of normal-ordered operators
as
\eqn\coo{ \hat N_{R,L}= N_{R,L} -a \ ,
\ \ \ \ \ \
\ \ a^{\rm (R)} =0\ , \ \ \ \ \ a^{\rm (NS) } =\ha \ .
}
Let us introduce the angular momentum operators $\hat J_1 \equiv \hat
J_{12}$ and $\hat J_2 \equiv \hat J_{34}$, which generate rotations
in
the respective 2-planes
(shifts in $\vp_1 $ and $\vp_2$ in \melv).
They can be written as
\eqn\eig{
\hat J_s=\hat J_{sL}+\hat J_{sR}\ ,\ \ \
\hat J_{sL}
= l_{sL} + \ha + S_{sL} \ , \ \ \
\hat J_{sR}
= - l_{sR} - \ha + S_{sR} \ , \ \ \ \ \ \ \ \ s=1,2 \ ,
}
where the orbital momenta in each plane
$l_{ L,R}=0,1,2,... $ are related
to the Landau quantum number $l$ and the radial quantum number $k$ by
$l=l_L-l_R$ and
$2k=l_L+l_R-|l|$,
and $S_{s R,L}$ are the spin components.
In the NS-NS sector, their possible values satisfy the condition
$$
|S_{1R}\pm S_{2R}| \leq \hat N_{R}+1\ ,\ \ \ \ \
|S_{1L}\pm S_{2L}| \leq \hat N_{L}+1\ .
$$
The mass spectrum is given by (cf. \magnetic)
$$
\a' M^2 = 2 ( \hat N_R+ \hat N_L ) + { \a' \ov R^{2} }
(m- b_1 R\hat J_1 -b_2R \hat J_2)^2
$$
\eqn\cosm{
+ \ { R^2w^2\over \a'}
- 2 \hat \g_1 (\hat J_{1R}-\hat J_{1L}) -2 \hat \g_2 (\hat
J_{2R}-\hat J_{2L})\ ,
}
where
\eqn\gam{
\N_R- \N_L = mw \ , \ \ \ \ \ \ \
\hat \g\equiv \g - [\g]\ ,\ \ \ \ \ \ \ \
\g _1 \equiv b_1 Rw \ , \ \ \ \g _2 \equiv b_2 Rw \ .
}
$[\g ]$ denotes the integer part of $ \g $ (so that $0\leq
\hat\g<1$).

In a generic magnetic background, one expects that all supersymmetries
will be broken since
fermions and bosons get different mass shifts:
the gyromagnetic
interaction is proportional to $b_s J_s$,
which is indeed different for fermions and bosons.
The supersymmetry in the case
of $b_1=b_2$ is due to a compensation
between the two gyromagnetic contributions
corresponding to the two 2-planes.
Let us consider, as an illustration, the fermion and boson states with
the winding number $w=0$,
K-K charge $m$, and
$\hat J^F_{1,2}=\ha , \ \hat J^B_1=1$, $\hat J^B_2=0$.
They have the same mass at zero
magnetic parameters
$b_1=b_2=0$, and there is a mass splitting for generic $b_1,b_2$,
proportional to $$ \delta M^2= (b_1-b_2) \big[ {m \over R} -\four
(3 b_1 + b_2)\big]\ . $$
It vanishes
in the supersymmetric case $b_1=b_2$.
As discussed above, the case $b_1=-b_2$ is also supersymmetric
(giving equivalent CFT related
by the redefinition
$\vp_2\to -\vp_2$). In this case, the superpartner of the fermion
with $\hat J^F_{1,2}=\ha$
is the boson
with $\hat J^B_1=-1$, $\hat J^B_2=0$.

\subsec{Partition function}

It is easy to compute the partition function in the GS
formulation following the discussion in \magnetic.
We first expand $y$ in eigenvalues of the Laplacian on
the
2-torus and redefine
the fields $z_1,z_1^*,z_2,z_2^*$ and $\psi_{R,L}^{+\pm }, \bar
\psi_{R,L}^{-\mp }$ to eliminate
the non-zero-mode part of $y$ from the $U(1)$ connection.
The zero-mode part of $y$ on the torus
($ds^2 = |d \s_1 + \t d \s_2 |^2 , \ \ \t=\t_1 + i\t_2 , \ \
0<\s_a\leq 1$) is $y_* =y_0 +
2\pi R(w \s_1 + w' \s_2)$, where $w,w'$ are
integer winding numbers. Integrating over the bosonic anf fermionic
fields
we get a ratio of determinants of scalar operators of the
type $\del + iA, \ \bd - i\bar A $ \ ($\del =\ha (\del_2 -\t
\del_1)$) with constant connections
\eqn\yyt{
A_s=b_s\del y_* = \pi \chi_s , \ \ \ \ \ \ \
\bar A_s= b_s \bd y_* = \pi \bar \chi_s \ , }
$$ \ \ \
\ \chi _s \equiv b_s R (w' -\t w) , \ \ \ \ \ \ \ \
\bar \chi_s \equiv b_s R ( w' -\bar \t w) \ ,\ \ \ \ s=1,2\ .
$$
The final expression for the partition function takes the following
simple form
  $$Z(R, b_1,b_2 ) = c V_5 R \int
{d^2\t \ov \tau_2^2 } \sum_{w,w'=-\infty}^{\infty}
\exp \big( - {\textstyle {\pi \ov \a' \t_2}} R^2
|w' -\t w|^2 \big)  \
{\cal
Z}_0 (\t, \bar \t;\chi_s,\bar \chi_s )  $$
\eqn\zzz{\times \
 \ {Y^2 \big(\t, \bar \t;
\ha (\chi_1+\chi_2 ) ,\ha (\bar \chi_1 +\bar\chi_2) \big) Y^2 \big(\t,
\bar \t;
\ha (\chi_1-\chi_2 ) ,\ha (\bar \chi_1 -\bar\chi_2) \big)\ov Y(\t, \bar
\t;
\chi_1 , \bar \chi_1 ) Y(\t, \bar \t;
\chi_2 , \bar \chi_2 ) }\ . }
Here
\eqn\yy{ Y (\t,\bar \t; \h, \bar \h)\equiv
{{\det}' (\del + i\pi \chi) \ {\det}' (\bd - i\pi \bar \chi)\ov
{\det}' \del \ {\det}' \bd } = {U(\t,\bar \t; \chi, \bar \h )\ov
U(\t,\bar \t; 0 , 0 ) } \ , }
\eqn\yyy{
U(\t,\bar \t; \chi, \bar \h )
\equiv \prod_{(n,n')\not=(0,0)}
(n'- \t n + \chi )(n'-\bar \t n + \bar \h )
\ , }
where, in the determinants, we have projected out the zero modes
appearing
at $\chi=\bar \chi=0$ (i.e. $Y (\t,\bar \t; 0, 0)=1$).
The equivalent form of $Y$ is
\eqn\ttyy{
Y (\t,\bar \t; \h, \bar \h) =
\exp[{{\pi (\chi-\bar \chi)^2
\ov 2 \t_2}}] \ {\theta_1(\h| \t)
\ov \h\theta'_1 (0| \t) } \ {\theta_1(\bar \h|\bar \t)
\ov \bar \h\theta'_1 (0|\bar \t) }
= \
\bigg|{ \theta \big[\matrix{\ha + bRw \cr \ha + bRw' \cr} \big]
(0| \t)
\ov bR (w' - \t w) \theta'_1 (0| \t) } \bigg|^2 \ , }
where $\theta_1(\h| \t) =
\theta \bigg[\matrix{\ha \cr \ha \cr }\bigg] (\h| \t)$.
In particular,
\eqn\ttyy{
Y \big(\t,\bar \t;\ha (\chi_1+\chi_2 )
,\ha (\bar \chi_1 +\bar\chi_2) \big)=
\
\bigg|{ \theta \big[\matrix{\ha + \ha (b_1+b_2) Rw \cr \ha + \ha (
b_1+b_2)Rw' \cr} \big]
(0| \t)
\ov \ha (b_1+b_2)R (w' - \t w) \theta'_1 (0| \t) } \bigg|^2
\ .
}
The factor ${\cal Z}_0$ in \zzz\ stands for
the contributions of the integrals over the constant parts of the
fields
$z_{1,2},\ z_{1,2}^* $ and $\psi_{R,L},\ \bar \psi _{R,L}$
(i.e. the contributions of $(n,n')=(0,0)$
terms in
the determinants)
\eqn\zerr{ {\cal Z}_0 = \tau_2^{-2} 2^{-8} R^4 |w'-\tau w|^4
{ (b_1+b_2)^4 (b_1-b_2)^4 \over b_1^2 b_2^2} \ .
}
Note that the full integrand of $Z$ is modular invariant
since the
transformation
of $\t$ can be combined with a redefinition of $w,w'$ (so that,
e.g.,
${\cal Z}_0$ and $Y$ remain invariant).

{}As expected, the partition function vanishes in the supersymmetric
limit
$b_1=b_2$ or $b_1=-b_2$.
More generally, due to the periodicity in $b_1R$ and $b_2R$,
\eqn\peri{
Z (R,b_1,b_2)= Z(R, b_1 + 2n_1 R^{-1}, b_2 +2n_2
R^{-1})\ , \ \ \ \ \ n_1, n_2 =0, \pm 1, ... \
, }
it vanishes also
at $b_1\pm b_2= 2n R^{-1} $ (these points are
zeroes of the theta-functions appearing in the numerator of \zzz).
The divergence at $b_{1,2} \to 0 $ (see eq.~\zerr ) corresponds to the
restoration
of translational invariance in the 1-2 and 3-4 planes
in the zero magnetic field limit. This divergence
reproduces the factors of areas of the 2-planes.
If both $b_1,b_2\to 0$, the divergence is cancelled against the fermion
factors in the numerators,
so that $Z$ also vanishes,
as it should in the supersymmetric zero-field limit.

 $Z$ is infrared-divergent for those values of the
parameters $b_1, b_2$ and $R$ for which there are tachyonic states
in the spectrum. This is seen by Poisson resummation in $w'$ and
expansion  of the integrand of \zzz\ at large $\tau _2 $ as in
\ourold . The integral over $\tau_2 $ in each term in the sum
diverges at large $\tau _2$ in the presence of tachyons, and  is
finite when tachyons are absent.

\subsec{Tachyonic states, T-dual background and
its possible evolution}

Let us discuss the tachyonic states in this model
for non-vanishing $b_1,b_2$.
Without loss of generality, we can assume $b_1\geq b_2>0$.
As in \magnetic, the state with lowest value for $M^2$ is a
particular spin 2 winding state with
$$
\hat N_{R}= \hat N_{L}=0\ , \ \ \ \
w=1\ ,\ \ \ m=0\ ,\ \ \ \ l_{1 L,R}= l_{2 L,R}=0\ ,
$$
\eqn\tacc{
S_{1R}= -S_{1L}=1\ ,\ \ \ \ S_{2R}=S_{2L}=0\ ,
}
so that $\hat J_1=\hat J_2=0$, $\hat J_{1R}-\hat J_{1L}=1$,
$\hat J_{2R}-\hat J_{2L}=-1$. Its mass (see \cosm)
\eqn\masss{
M^2= {R^2 \over \a'^2 }- {2R\ov \a'} (b_1-b_2)
\ }
becomes tachyonic for
\eqn\kokp{
b_1-b_2> {R\over 2\a'} \ . }
The negative contribution to $M^2$ originates from the gyromagnetic
interactions in \cosm \ depending on non-zero
winding number.\foot{As follows from the
discussion of the solution of the tachyon equation in
\refs{\ourold,\magnetic,\RTN}
the tachyon wave function in the $N=1$ Melvin model
is localized
(with finite width related to $\g=bRw$)
near $r=0$. The same will be true for $N >1$ models.
This is similar to localization of tachyons
at fixed planes in orbifold models.
Another analogy is that here the
tachyonic states must have non-zero $S^1$ winding number
{\it and}
spin in a 2-plane while
the tachyons in orbifold models appear in the
twisted sector.}

The theory with $b_1=b_2$ (or $b_1=-b_2$)
is tachyon-free for any radius
$R$. This is in agreement with the above discussion
implying the presence of partial unbroken supersymmetry
for $b_1=\pm b_2$.


The non-supersymmetric string models with
$R\geq 2\a '(b_1- b_2)$
are also tachyon-free. It is easy to show that the
integral over $\tau_2$ in the one-loop
partition function \zzz\ is finite  in this case, for
each term in the sum.

An important question is how the string background
reacts to
the presence of tachyons.
To address this problem, it is useful to consider the
T-dual string model, where the tachyon has the same
quantum numbers as in \tacc ,
except for interchanged winding and momentum numbers, i.e.
$w=0$ and $m=1$.
Since the tachyon \tacc\ now appears not in the winding but
in the {\it momentum}
sector, it is a state of the
supergravity multiplet.
That means that, as in the
1-parameter case discussed in
\RTN, here the instability can be seen directly at the
level of the supergravity equations expanded near the
T-dual background. Above some critical radius,
the space-time background becomes unstable under
the tachyonic perturbations of the supergravity fields.

Applying T-duality in the $y=x_9$ direction
to \melv\ gives the following T-dual background
(string-frame metric, dilaton, and NS-NS 2-form)\foot{Note that
this solution has regular curvature and the
dilaton can be made small everywhere.}
\eqn\melvp{
ds^2_{10} =-dt^2+dx_i^2+dr_1^2+dr_2^2+ r_1^2 d\vp_1^2+ r^2_2 d\vp_2^2
+ f^{-1} \big[ dy^2
- (\td b_1 r_1^2 d\vp_1 + \td b_2 r_2^2 d\vp_2 )^2\big]\ ,
}
\eqn\molp{
e^{2(\phi -\phi_0)}=f^{-1}\ ,\ \ \ \ \ \ \ \ \ \
B_2 = f^{-1}\big( \td b_1 r^2_1
d\vp_1
+\td b_2 r^2_2 d\vp_2\big)\wedge dy
\ ,\
}
$$
f\equiv 1+\td b_1^2 r_1^2+\td b_2^2 r_2^2\ .
$$
Here, for convenience, we have renamed $(b_1,b_2)\to (\td
b_1,\td b_2)$.
The counterpart of the state \masss \ has the mass
($R\to \td R = { \a'\ov R}$)
\eqn\maxxx{
M^2= {1 \over R^2}-{2 \over R} (\td b_1- \td b_2)\
\ .
}
Thus this background is unstable for $R>R_{\rm cr}$,\ \
$R_{\rm cr}=\big( 2\td b_1-2\td b_2\big)^{-1}$.

Consider now the case in which the magnetic field is
$\td b_1=\td b_2+ {1\ov 2R } +\epsilon ,\ \epsilon>0$.
For small $\epsilon $, the only tachyonic state
is the first Kaluza-Klein mode with $m=1$
and mass \maxxx.
The corresponding
perturbation of the background \melvp,\molp ,
as a solution of the linearized supergravity equations,
will then grow exponentially with time.
The explicit form of this tachyonic perturbation
mode gives
an indication (at least to linear order)
of the
evolution of the geometry from the unstable state.
Having $m=1$ and vanishing orbital angular momenta
(see \tacc), the
spatial dependence of the unstable mode
is of
the form $
f(r_1,r_2) \cos{y\ov R}$.
Since the tachyonic state has
$S_{1R}, S_{1L}\neq 0$, the perturbation
will modify
the metric and the antisymmetric tensor
in their parts corresponding to the 2-plane $(r_1, \vp _1)$.

This suggests that
the background \melvp,\molp\ will be evolving
towards a new solution which is {\it not}
translational invariant in $y=x_9$.
There is
some similarity with the fate of the well-known
instability
\GLF\
of the black string solution. The latter background
(which, like \melvp, has translational isometry in $x_9$)
becomes unstable
for some $R > R_0$, and seems to evolve
into a configuration which is no longer translationally
invariant in $x_9$ \horo .

To try to check this conjecture, one is to generalize
\melvp,\molp\
to include time dependence (in particular, making the effective
$\td b_s$ parameters ``run'' with time).
It would be interesting to find the explicit solution for the
end-point of the evolution.
The resulting background should be stable at
the supergravity level, but
may still be unstable at the full
quantum string theory level.

In general, non-supersymmetric string models that are stable
(non-tachyonic) at the classical level may become unstable at the
quantum string level. For example, the tachyon-free string
background with $R\geq 2\a '(b_1- b_2)$, \ $b_1\neq \pm b_2$ may
decay due to quantum-mechanical effects. Since the physical
mechanism for the decay here is different from the one discussed
above the final state of the evolution may also be different. If
initially $b_1$ is very close to $b_2$, then it is natural to
expect that this model will evolve towards a
(quantum-mechanically) stable supersymmetric model with $b_1'=b_2'
< {\rm min}(b_1,b_2)$ (see related discussion in section 4).
\foot{In the special case of rational $b_s R= 1/n_s $, when the
model becomes similar to the orbifold model, this is analogous to
the picture suggested in \ada: a non-supersymmetric $n_1 \not=\pm
n_2$ unstable C$^2/Z_m$ orbifold should decay into the
supersymmetric one (see also \TTU ).}

\subsec{Four-parameter  generalization}

As in the $N=1$ case \magnetic, the
T-dual string model based on \melvp,\molp\
admits a straightforward generalization:
introducing two extra parameters
into \melvp , \molp\ by the
locally trivial transformation
$\vp_s \to \vp_s + b_s x_9$
we get a string model that depends on the four magnetic
parameters
$(b_1,b_2,\td b_1,\td b_2)$ (as well as $R$).
The corresponding \sm\ Lagrangian is then given by
$$
L= \del_+ x_i \del_- x_i +  \del_+ r_1
\del_- r_1
+ \del_+ r_2 \del_- r_2+ r_1^2 \del_+ \vp_1' \del_- \vp_1'
+ r_2^2 \del_+ \vp_2 ' \del_- \vp_2'
$$
\eqn\leagg{
+ \ {f\inv} \big[ \del_+ y \del_- y -
(\td b_1 r_1^2 \del_+ \vp_1' + \td b_2 r_2^2 \del_+ \vp_2')
(\td b_1 r_1^2 \del_- \vp_1' + \td b_2 r_2^2 \del_- \vp_2')
\big]
}
$$
+\ {f\inv } \big[ \td b_1r_1^2( \del_+ \vp_1' \del_-y -
\del_-
\vp_1' \del_+y)
+ \td b_2r_2^2( \del_+ \vp_2' \del_-y - \del_- \vp_2' \del_+y)
\big]
+ {\cal R} (\phi_0 - \ha \ln f)\ ,
$$
$$ \vp_{1}'\equiv\vp_{1}+b_{1}y \ , \ \ \ \ \ \ \ \
\vp_{2}'\equiv\vp_{2}+b_{2}y \ , \ \ \ \ \ \ \ y=x_9\equiv x_9 + 2 \pi
R \ .
$$
This four-parameter model
is exactly solvable,
despite  its
complicated curved-space geometry.
The computation of its spectrum
can be done by a simple
generalization
of the case $b_2=\td b_2=0$ considered in \magnetic\
and the $\td b_1=\td b_2=0$ case in \cosm--\gam.
One finds (cf. \cosm)
$$
\a' M^2 = 2 ( \hat N_R+ \hat N_L ) + { \a'\ov R^2}
(m- b_1 R\hat J_1 -b_2R \hat J_2)^2
$$
\eqn\sepee{
+ \ { \a'\over \td R^2} (w - \td b_1 \td R \hat J_1 - \td b_2 \td
R
\hat J_2)^2
- 2 \hat \g_1 (\hat J_{1R}-\hat J_{1L}) -2 \hat \g_2 (\hat
J_{2R}-\hat J_{2L})\ ,
}
$$
\N_R- \N_L = mw \ , \ \ \ \ \ \ \ \
\td R={\a'\over R} \ ,\ \ \ \ \ \hat \g_s\equiv \g_s - [\g_s]\ ,\
$$
$$
\g _1 \equiv b_1 Rw + \td b_1 \td R m
- \a' \td b_1  b_s \hat J_s \ , \ \ \ \ \ \ \ \
\g _2 \equiv b_2 Rw + \td b_2 \td
R m
- \a' \td b_2  b_s \hat J_s \ .
$$
It is possible also to compute the partition function,
following the
methods of \magnetic.\foot{This was
recently done in \TTU\ which appeared after
the completion of the present
work.}
 A close inspection of the spectrum shows
that it is supersymmetric in the case  when
$b_1= b_2$, \
$\td b_1=\td b_2$ (or $ b_1= - b_2$, \ $\td b_1=- \td b_2$).
Special cases are $\td b_1=\td b_2=0$
or   $b_1= b_2=0$  when this model
reduces to
the $N=2$ Melvin model
with $b_1=\pm b_2$ or  to its T-dual \melvp , \molp\
with $\td b_1=\pm \td b_2$.

\newsec{ Supersymmetric R-R vector fluxbrane backgrounds }

\subsec{Supersymmetry}

Starting with the background \melpp\ (or \melv\ and its T-dual
\melvp,\molp) we may now construct
other related solutions by applying U-duality.
An equivalent procedure
is to start with the same flat background in $d=11$, with
the KK coordinate
$y$ now interpreted as $x_{11}$, and obtain
type IIA magnetic R-R flux tube backgrounds by reducing
along $x_{11}$.
This will give an $N$-parameter generalization of the
R-R flux 7-brane \refs{\DG,\green}.

Let us first discuss the issue of supersymmetry of the resulting
solutions on the example of the 2-parameter case.
The background described by the locally flat metric \melv\
is obviously 1/2 supersymmetric for $b_1=b_2$
as a solution of 11-d supergravity:
the condition of existence of Killing spinors in $d=11$
is the same
\kii--\fef\ as in $d=10$.
The supersymmetry should of course be
preserved at the full M-theory (membrane theory) level,
but it might be broken by the 11 $\to$ 10 reduction
if one restricts consideration
to the 10-d supergravity (or perturbative 10-d string theory) states
only.
The necessary condition for
preserving the supersymmetry at the supergravity level
is that Killing spinors should be constant
along the direction of
compactification.\foot{More precisely,
the number of supersymmetries preserved upon reduction of a
supersymmetric solution
is equal to the number of Killing spinors which have vanishing Lie
derivative along the direction of reduction (see
\fip\ and references there
for a discussion and examples).}
In other words, the supercharges that are preserved by the reduction
form a subset of
those of the original solution that commute
with translations along the KK direction.

As discussed in section 2, in the case of $b_1=b_2$ in $N=2$ model
(and in similar $N >2$ cases like \sii)
the Killing spinors are actually
constant, i.e. do not depend on $y$.
That means that all of the 16 supersymmetries of the 11-d solution
must be present in the resulting type IIA solution.\foot{It
should be straightforward to
construct the GS string action in the corresponding
R-R background in IIA
theory by using the same method as in \green,
i.e. by starting with the corresponding
supermembrane action in flat 11-d background.
The existence of a residual global supersymmetry of
the resulting GS action would then imply supersymmetry and thus
stability of the magnetic R-R fluxbrane background.}
{}From the 10-d supergravity perspective,
the supersymmetry condition $b_1=\pm b_2$
should arise from solving the type IIA Killing spinor
equation in the
corresponding curved R-R background.

This is no longer so in other possible 11-d supersymmetric
cases (like \peerr) where
there is a relation between the $b_s$-parameters and $R$:
there the Killing spinor depends on $y$ and the
supersymmetry will be broken
at the supergravity level by the dimensional reduction.
The simplest example is provided by the special
1-parameter case
$b_1 R=2$, $b_2=0$ \green.
Here the 11-d space is globally flat ($\vp_1'=\vp_1+ b y$
is a globally defined $2\pi$-periodic
coordinate), i.e. the 11-d theory is equivalent
to the
flat M-theory with 32 supersymmetries.
However, the resulting
10-d background
is non-trivial and non-supersymmetric
(both at the level of type IIA
supergravity and the perturbative string
theory in the corresponding R-R background \green).
It is only after the non-perturbative (wound membrane)
M-theory states are taken into account, this theory
should regain supersymmetry
and become equivalent to the standard maximally
supersymmetric type IIA string or M-theory
in flat background.

\subsec{General form of the solutions}

Let us start with the $d=11$ analog
of \melpp, i.e. a $d=11$ Minkowski spacetime
with twists in $N=1,2,3,4$ orthogonal spatial planes
and $y\equiv x_{11}$
($n=2N+1, ..., 9$, \ $s=1,..., N$)
\eqn\melvv{
ds^2_{11}=-dt^2+dx_n^2+dx_{11}^2+\sum^N_{s=1} [dr_s^2+ r_s^2
(d\varphi_s + b_s
dx_{11}) ^2] \ .
}
Being locally flat, this is an exact solution of 11-d
supergravity and also of M-theory.
Dimensional reduction in the $x_{11}$ direction gives
the following string-frame type IIA metric, dilaton $\phi$
and R-R vector 1-form $A$ \refs{\DG, \Ur }
\eqn\rrmv{
ds^2_{10A}=f^{1/2}\bigg[-dt^2+dx_n^2+\sum^N_{s=1}( dr_s^2 +r_s^2
d\vp_s^2 )
- f^{-1} ( \sum^N_{s=1}b_s r_s^2 d\vp_s)^2 \bigg]
}
\eqn\sss{
e^{2(\phi -\phi_0)}=f^{3/2}\ ,\ \ \ \ \ \ \ A = f^{-1} \sum^N_{s=1} b_s
r^2_s
d\vp_s
\ ,\ \ \ \ \ \ \ f=1+\sum^N_{s=1} b_s^2 r_s^2 \ .
}
This background may be interpreted
as representing an orthogonal intersections of $N$
individual R-R F7-branes \refs{\DG,\green}.

A supersymmetric solution preserving $1\over 2^{N-1}$ \ ($N >1$)
fraction of
maximal supersymmetry (i.e. having $2^{6-N}$ real supersymmetries)
is obtained by demanding \foot{As already mentioned above,
choices of other
signs of $b_s$ give equivalent solutions
related by $\vp_s\to -\vp_s $.}
$$b_1= b_2+...+b_N \ . $$
This gives, in particular,
a two-parameter family of three intersecting F7-branes,
or a three-parameter family of four intersecting
F7-branes.\foot{One may also consider the case of
$N=5$  with $x_{11}=y$
 (and one of the angles playing the role  of
Euclidean time).  The corresponding type IIA
background may be interpreted as ``F(-1) brane''.}

The simplest non-trivial case of \rrmv,\sss\ is
$N=2$, or explicitly
\eqn\lv{
ds^2_{10A}=f^{1/2}(-dt^2+dx_n^2+dr_1^2+dr_2^2+r_1^2 d\vp_1^2+
r^2_2 d\vp_2^2)
- f^{-1/2} ( b_1 r_1^2 d\vp_1+b_2 r_2^2 d\vp_2 )^2 \ ,
}
\eqn\defo{
e^{2(\phi -\phi_0)}=f^{3/2}\ ,\ \ \ \ \ A = f^{-1}\big( b_1 r^2_1
d\vp_1+ b_2 r^2_2 d\vp_2 \big)
\ ,\ \ \ \ \ \ f=1+b_1^2 r_1^2+b_2^2 r_2^2\ .
}
This 10-d background
may be interpreted as describing
an orthogonal intersection of the two R-R flux 7-branes
(as expected for a localized intersection of two branes,
this solution does not have radial symmetry in 4
common transverse directions).
The
10-d description is valid for sufficiently
small values of $r_1,r_2$ and $b_1 R,b_2R$.
This solution is related to the NS-NS type IIA
background \melv\ (or its $T$-dual \melvp,\molp)
by U-duality, i.e. by the $T_9ST_9$ sequence of duality
transformations (which produces the ``9-11 flip").

As discussed above, since the Killing spinor
corresponding to \melv\ is constant in the
$b_1=b_2$ case,
the supersymmetry is preserved by the duality,
so the dual $b_1=b_2$ R-R background is also 1/2 supersymmetric.
Its explicit form is
\eqn\melvo{
ds^2=f^{1/2}\big( -dt^2+dx_n^2+dr_1^2+dr_2^2
+r_1^2 d\vp_1^2+ r^2_2 d\vp_2^2\big)
- b^2 f^{-1/2} ( r_1^2 d\vp_1+ r_2^2 d\vp_2 )^2\ ,
}
\eqn\tyu{
e^{2(\phi -\phi_0)}=f^{3/2}\ ,\ \ \ \ \ \ \ \ A = b f^{-1}\big(
r^2_1 d\vp_1+ r^2_2 d\vp_2 \big)
\ ,\ \ \ \ \ \ \ \ f=1+b^2 (r_1^2+ r_2^2) \ . }
This solution is, in fact, equivalent
to the supersymmetric
``F5-brane'' of \GS: the two backgrounds
are related by the following coordinate
transformation:
\eqn\coop{r_1 = r \cos \theta\ , \ \ \ \ r_2 = r \sin \theta\ , \ \ \ \
\vp_1 =\td \phi + \psi\ , \ \ \ \ \vp_2 =\td \phi - \psi \ . }
Since this solution has
$F_2$ fluxes through two orthogonal planes
 (rather than  $F_4$ flux)
its more appropriate
interpretation is that of two intersecting R-R F7-branes
of equal fluxes.

By applying $T_9ST_9$ duality to the 4-parameter background in \leagg
\ (or to its obvious $2N$ parameter generalization) one obtains
more general fluxbrane intersections which,
in addition to the R-R 1-form potential,
have non-trivial NS-NS 2-form field.

\subsec{Classical and quantum (non-perturbative) instabilities}

Let us now discuss
stability of  non-supersymmetric F7-brane intersections,
and possible fate of the unstable backgrounds.
In order to make use of the discussion of the dual NS-NS model,
let us assume that $x_9$ (one of the $x_n$ coordinates in
\rrmv)  is periodic  with radius $R_9$,
so that the R-R \rrmv,\sss\ and the NS-NS
\melpp\ backgrounds  may be related through  the ``9-11'' flip
in 11 dimensions.
The presence and nature of instabilities
(classical or quantum)
as well as the final state of evolution of an unstable background
may depend on the values of the string coupling
(or $R_{11}$)
 and the
radius $R_9$.
For example, the NS-NS background \melv\ is stable classically
as a supergravity solution for large $R_9$, but
for $R_9 <  \sqrt{\a'} $ the theory is more
appropriately  represented
by the T-dual background \melvp,\molp, which is
classically unstable  (for
certain values of the magnetic field parameters, see  section 3.4)
as a supergravity solution.
In what follows, we will assume that the string coupling is small,
 i.e.
$R_{11}\ll l_P$.

Let us first consider the issue of  classical instabilities
 of the
above R-R backgrounds
at the supergravity level.
A supergravity solution is stable if there are no
tachyonic modes in the field  equations
expanded near the corresponding background.
The NS-NS  background \melpp\
is stable at the supergravity level, in both
$d=10$ and $d=11$ theories.
In $d=10$ this  can be seen  directly from the
spectrum of the NS-NS string model in section 3
which contains no tachyons in the zero-winding sector, i.e.
(as in the $N=1$ case \refs{\magnetic,\RTN})
there are no tachyonic states among
the supergravity fluctuation modes for
any value of the magnetic field.
As a result,  the F7-brane intersections that arise
by the $11\to 10$ dimensional
reduction must also be stable
(the 10-d supergravity modes are a subset of
the 11-d supergravity modes,
none of which is tachyonic).

Note that the fact that, e.g.,  the $N=2$  R-R solution
 \lv,\defo\
is U-dual to \melv\ (which
is stable at the supergravity level) is not a priori
 sufficient to argue for the stability of the former
 background.\foot{
For example, \lv\ is  U-dual also to
the background \melvp\ which, for a certain values  of parameters,
is unstable at the  $d=10$  supergravity level.
Here the instability is developing in the compact isometry
 direction
and thus is ``overlooked'' by the T-(or U-) duality argument
(though $T-duality$ is a symmetry of the supergravity equations, it
applies only in the presence of an isometry,
while fluctuations may depend on the isometry direction).
}
Instead, we use that since
 \melv\ is stable as a 10-d supergravity solution, its
{\it direct} lift to 11 dimensions \melvv\ should  also be a stable
11-d supergravity background.\foot{
By the same logic, its reduction to 9 dimensions
should be giving a stable 9-d solution.
Indeed, it leads to
the KK Melvin 9-d  background which is known to be stable at the
gravity level.}
The R-R background is then a different 10-d reduction of  the same
stable 11-d  background.

There is, however,
a potential tachyonic instability of the flat 11-d space \melvv\
with
$b_1\not=b_2+ ...+ b_N $
at the full  M-theory level.
 Indeed, we know that
the  subsector of M-theory
representing the  perturbative type IIA strings in the
10-d background \melpp\ is unstable in the winding string
sector for certain values
of the magnetic  parameters.
{}From the 11-d perspective, the instability is
due to  special  winding membrane states.
For example, in the $N=2$ case such  state is a  counterpart
of the winding string mode \tacc.
Written in terms of 11-d parameters, its mass \masss\ is
\eqn\qqqy{
M^2=(4\pi^2 w R_9 R_{11} T_2)^2 - 8\pi^2 R_9 R_{11} T_2 (b_1-b_2)\ ,
}
where $T_2=(2\pi l_P^3)^{-1}$ is the membrane tension.
This means that type IIA string theory in the R-R
background \lv,\defo\ should also be unstable at a
non-perturbative level
in a certain range of the parameters $b_1-b_2$, $R_9$, $R_{11}$
(eq.\qqqy\ is applicable for $R_9\ll l_P$ or $R_{11}\ll l_P$).

In the case of the direct 11-d
lift of the T-dual background \melvp,\molp\
the counterpart of the winding membrane  state is
a mode of the supergravity multiplet.
In this case the presence of
a  tachyonic instability
(and  possible evolution of the unstable background)
can be studied by  using  simply  the 11-d supergravity
equations (see also \RTN).

\bigskip

Let us now comment on possible
quantum instabilities.
The physical phenomenon leading to  quantum
decay of F7-brane configurations is
the KK monopole (D6-brane) pair production in the magnetic field
\refs{\WWW,\gau,\gauu}.
For the single  magnetic field parameter $b_s=b$ ($N=1$),
the semiclassical instanton amplitude was discussed  in
\refs{\gauu ,\CG} , and  is of order
$e^{- I }$,  where  $ I \sim  V_6  \kappa_{10}^{-2}  R b^{-1} $.
In the case of  intersection of $N$ F7-branes
with generic parameters $b_s$ \DG\
the production of monopoles should be at the
expense of the magnetic energy density
proportional to $\sum_s b_s^2$.
It is natural to expect that due to
pair creation the values of
$b_s$ should decrease (in general, at different rates),
and the process should not stop until
the background ``rolls down'' to one of the supersymmetric
configurations with $\sum_s \pm b_s =0$.
When that happens, the instanton amplitude should vanish
identically due to
the presence of  fermionic zero modes
in the supersymmetric background.\foot{Since
the presence of supersymmetry
for special values of $b_s$ was not noticed in \DG,
it was assumed that these intersecting fluxbrane backgrounds should
always decay  non-perturbatively.}

As was mentioned in section 3, which particular supersymmetric
vacuum is reached should depend on  initial
values of the magnetic parameters $b_s$.
For example, the initial
configuration with $b_1=b_2+\epsilon $,
$|\epsilon|\ll 1$ should evolve  into the
one with $b_1=b_2$, while the configuration
with $b_1=\epsilon $, $b_2\gg b_1$
may roll down to the trivial $b_1=b_2=0$ vacuum.


\newsec{Concluding remarks}

We have seen that there is a novel type of
R-R brane backgrounds preserving
fractions of supersymmetries which are closely related
(U-dual)
to simple solvable NS-NS string models
with continuous parameters.
One interesting application of these backgrounds
 is to the
study of the evolution of instabilities
in closed string theory.
As emphasized in the previous sections, in some related  (T-dual,
section 3.4)
models the tachyons appear
at the supergravity level, so the problem of  decay of
due to closed string tachyons may be effectively
addressed by solving time-dependent
supergravity equations of
motion.\foot{Here the  evolution is  caused by
unstable fluctuation  modes in the supergravity
multiplet itself. This is different from the case
discussed in  \ada.}
We have argued that non-supersymmetric, but tachyon-free,  backgrounds
should decay
into  stable supersymmetric ones  with
$\sum_{s=1}^N \pm b_s =0$ (including, in particular,
 the trivial case of  all $b_s=0$).

There are also other potential applications
which are worth of further study.
The existence of the supersymmetric F1, F3, F5 branes
representing orthogonal intersections of 4, 3 and 2
F7-branes  which have, respectively,
4,8 and 16 supersymmetries, may provide a new tool
for ``brane-world'' model building.
The dual NS-NS string models are
also of interest in this context,  being,  in some respects,
continuous-parameter
analogs of non-compact orbifolds.
 The breaking of supersymmetry, and thus the
  splitting between fermion and boson masses
  in ``parallel'' dimensions
here is   controlled by  a number of  continuous parameters $b_s$.
Putting  standard D-branes in these
NS-NS backgrounds  may  lead  to  new examples
 (in addition to branes on orbifolds and
 conifolds) of
 supergravity duals  of
  non-conformal
 gauge theories with reduced
 amount of supersymmetry.
 We discuss  some relevant solutions
 in Appendix.

An  interesting   problem is to understand the structure of
the
world-volume theory, corresponding,
 in particular,  to the
 most interesting  case  of the R-R  F3-brane.
 While the standard  supersymmetric p-branes
 (supported by delta-function sources  that  can be put
  anywhere in transverse space)
 are parametrized by harmonic functions
  decaying at infinity, have  finite
 mass (or charge) density  and can be BPS-superposed,
 this is not so for the above  R-R  Fp-branes.
 The latter are, in fact, more   analogous to   fractional
 Dp-branes  or D(p+q)-branes wrapped on $q$-cycles
 which are supported by smooth fluxes instead of
 delta-functions and  are localized in transverse
 space (cf. \ktmn; see also \Ur\ for related discussion).

One may  argue that the breaking of 3/4 of
supersymmetry  should
imply  the presence of  24 fermionic collective coordinates
in   the 1+3 dimensional world-volume theory.
 Since  the  F3-brane brane cannot move freely in
 the  transverse 6-space
 (being the intersection of the three  F7-branes it is
  pinned down at $r_1=r_2=r_3=0$)
 the translational symmetry breaking
 should  not produce
 goldstone bosons,
 but the breaking  $SO(6)\to U(1)^3$
 of the  rotational  symmetry
 in transverse directions  should  give
15-3=12  massless scalar  bosons on the brane.
The expected  $\cal N$=2, $d=4$  world-volume
supersymmetry (corresponding to 8  preserved supercharges)
can be realized by arranging
 these  degrees of freedom
into the 3 hypermultiplets (the numbers of on-shell
modes do   match: $\ha \times 24 = 12$),
in agreement with  a discrete   symmetry
interchanging the 3  planes.\foot{One may use instead of \sii\
a  more symmetric form of the  supersymmetry condition
$b_1 + b_2 + b_3=0$ to  make this $Z_3$ symmetry manifest.}
The  matching between bosons and fermions is
 also consistent  with
the expectation that there should be no  additional
vector gauge bosons on the F3-brane
(the only non-trivial  R-R flux  is  associated with
the 1-form field  having scalar gauge parameter).\foot{Note also
that   there is no quantized number
    that  could  be associated
  with a rank of a non-abelian gauge group --
 the parameters $b_s$ are continuous.}


While the  standard  or fractional  Dp-branes
have    ``dual''  open-string  description
in flat space (which  gives an alternative  way  to
determine the structure of
massless modes on the
brane)  an   existence  of a similar
description  for the supersymmetric Fp-branes
remains an open question.
In general,
it  would be interesting  to study
the  open superstring  spectrum
and possible D-brane configurations
in the flat NS-NS  backgrounds  \melpp.
For example, one can  place a Dp-brane along parallel
directions $x_i$.
The  corresponding open-string CFT is, in principle,
 straightforward to
solve explicitly.
 The  corresponding supergravity solution is also readily
constructed and is  discussed   in  Appendix.

\bigskip

\noindent
{\bf Acknowledgements}

\noindent
We are grateful to S. Frolov,
G. Horowitz, I. Klebanov,
H. Liu, R. Myers, H. Nieder,
E. Poppitz and A. Uranga for a useful discussions.
J. R. was supported by Universidad de Buenos Aires and Conicet.
The work of A.A.T. is partially supported by the DOE grant
DE-FG02-91ER40690,
PPARC SPG grant 00613, INTAS
project 991590 and
CRDF Award RPI-2108.

\appendix{A}{D3 branes with reduced supersymmetry}
\def\bR{{\rm R}}

Here we construct D3 branes with reduced supersymmetry
by adding magnetic
 fluxes in several planes of the transverse 6-space
 as in \melpp.\foot{Solutions of this type  in the $N=1$ case where
there is no  supersymmetry
were discussed  in
\RTT.}

As a first example, consider a D3 brane with transverse space $\bR
^2\times \bR ^2\times \bR^2$. Suppose one of the 3 ``parallel''
 coordinates
$x_3=y$ is compact with period $2\pi R$.
Starting with the standard D3-brane solution and
making formal coordinate redefinition
$ \vp_s\to \vp_s+b_s y$ which  mixes $y$ with
 angles in  the three transverse planes,
we get  a  new, globally-inequivalent,
 type IIB solution with
 $F_5$ having the standard self-dual form,
 dilaton being constant, and  the following metric
\eqn\dddt{
ds^2_{10B}=H^{-1/2}\big[-dt^2+dx_1^2+dx_2^2+dy^2\big] + H^{1/2}
\sum_{s=1}^3 \big[
dr_s^2+r_s^2 (d\vp_s+b_s dy)^2 \big] \ .
}
Here  $H\equiv h(r_1,r_2,r_3,\vp_1+b_1 y,\vp_2+b_2 y  ,
\vp_3+b_3 y)$, where $h( r_1,r_2,r_3, \vp_1,\vp_2,\vp_3)$
is the original
 harmonic function on  the transverse  $\bR^6$ space.
 For example, if one starts with the
   standard spherically-symmetric D3-brane, then
   \eqn\loo{H=h=h_0 + { L^4\ov (r^2_1 + r^2_2 + r^2_2)^2 } \ . }
Alternatively, one may    choose, for example,
 $H=h=\sum_{s=1}^3  a_s \log {r_s\ov r_{0s}} $.
If the theory in the absence of D3-brane ($H=1$)
 is supersymmetric, i.e. $b_s$
satisfy \sii, then this solution preserves 1/8 of maximal type IIB
supersymmetry. Indeed, the conditions \koo\ viewed as restrictions on
a 6-d spinor, i.e.
$
(I- \g_{4567}) \ep_0 = 0 ,
(I- \g_{4589}) \ep_0 = 0 $
 should be supplemented by the D3-brane  condition \duff\
$ (I-i\gamma_{456789})\ep_0 =0$.\foot{
Here we use the indices  4,...,9 for the transverse directions,
 $\g_k$ are 6-d Dirac matrices, and $\ep_0$ is the 6-d part
 of the constant factor of the Killing spinor.}
   These are  equivalent to
$\g_{45} \ep_0 =  i \ep_0 , \
\g_{67} \ep_0 = - i \ep_0 ,\
\g_{89} \ep_0 = - i \ep_0 ,$
implying  that 1/8 of maximal supersymmetry is preserved.
This suggests  that the corresponding world-volume
3-d gauge theory (assuming we compactify in $y$)
should have four unbroken supercharges.

 Consider  the  choice of $h_0=0$, i.e. $H= { L^4\ov (r^2_1 + r^2_2 +
r^2_2)^2}$. In the UV region where all $r_i$ are large the metric
will factorize into $AdS_5$ with its  $S^1$ direction $y$ being
``mixed''   with $\vp_s$ coordinates of $S^5$.\foot{We are
 grateful to I. Klebanov for suggesting
this possibility and pointing out a relation to the work of \gan.}
One thus obtains a supergravity solution which should be dual to
SYM theory with 0, 4, or 8 unbroken supersymmetries, depending
 on the values of $b_s$.
 The gauge
theory (IR RG flow) interpretation of this background remains to
be investigated.

A different  D3-brane  solution can be obtained by
starting with the standard  D2-brane solution  with the
  transverse
7-space being flat ``twisted'' $S^1 \times \bR^2 \times \bR^2
\times \bR^2$ space (a similar D4-brane solution is given
explicitly below). If the corresponding harmonic function  $H$ is
chosen to be $\vp_s$-independent, then one may do T-duality in $y$
(cf. \melvp,\molp), obtaining a D3-brane background  which is
similar to \dddt\ at small $r_s$. In general, it  will have an
extra factor of $f^{-1}$, $f\equiv 1 + b^2_1 r^2_1 + b^2_2 r^2_2 +
b^2_3 r^2_3 $ in front of $dy^2$, with extra $B_2$ field and the
dilaton $e^{2 \phi-2 \phi_0} = f^{-1} $ as in \molp. This is a
particular case of the solutions obtained in \gan .

One may also  consider a  D3-brane solution
 with the  transverse   6-space
being a product of a line $\bR$ (or a circle) with  the  $N=2$
twisted version of $\bR ^2\times \bR ^2\times S^1$. The
corresponding $d=4$  gauge theory on the brane will have $\cal
N$=2 supersymmetry. Even though the dilaton here is constant, for
a logarithmic choice of   the corresponding harmonic function,
this background may  be  reflecting (as  in the orbifold
examples in \oorb) the running  of the $\cal N$=2 gauge theory
coupling with scale.

 The T-dual to this D3-brane
  is a   D4-brane solution  with
 twisted $\bR^2\times \bR ^2\times S^1$ as the 5-d
transverse space. It has the form \eqn\ddda{
ds^2_{10A}=H^{-1/2}\big[-dt^2+dx_i^2\big] +
H^{1/2}\big[\sum^2_{s=1} \big( dr_s^2+r_s^2 (d\vp_s+b_s dy)^2\big)
+dy^2 \big] \ , }
$$
e^{2\phi} =H^{-1/2}\ ,\ \ \  F_4 = *(d H^{-1}
\wedge dx_0\wedge ...\wedge dx_4) \  , \
$$  where $
 H=H$ is a  harmonic function in  $\bR^4$
 with $\vp_s \to \vp_s+b_s y$, 
 in particular, $H=H(r_1,r_2)$. 
For $b_1=b_2$  the corresponding world-volume theory should have
$\cal N$=1, $d=5$ supersymmetry.

Analogous  solutions can be also obtained by
starting with the 11-d M5-brane background
 with the transverse 5-space
$\bR
^2\times \bR ^2\times \bR$ being mixed  with a compact ``parallel''
coordinate $y$ ($i=1,...,4$)
\eqn\mmmt{
ds^2_{11}=H^{-1/3}\big[ -dt^2+dx_i^2+dy^2\big]+ H^{2/3}
\big[ dr_1^2+r_1^2 (d\vp_1+b_1 dy) ^2+dr_2^2+r_2^2 (d\vp_2+b_2 dy) ^2+
dz^2\big]
}
Here
$
H=H(r_1,r_2)$ is again the corresponding harmonic function.
Dimensional reduction in $y$ gives a generalized
type IIA
D4-brane solution with a R-R $F_2$
flux in the two transverse planes
$$
ds^2_{10A}=f^{1/2} \bigg(H^{-1/2}\big[ -dt^2+dx_i^2\big]+ H^{1/2}
\big[ dr_1^2+r_1^2 d\vp_1 ^2+dr_2^2+r_2^2 d\vp_2 ^2+ dz^2
$$
\eqn\rrry{
-\ H f^{-1} (b_1r_1^2 d\vp_1 + b_2 r_2^2 d\vp_2 )^2\big]\bigg)\ ,
}
$$
e^{2\phi }=f^{3/2}H^{-1/2}\ ,\ \ \ A_1=Hf^{-1}(b_1r_1^2 d\vp_1+b_2
r_2^2
d\vp_2)\ ,
\ \ \ f=1+H(b_1^2r_1^2+b_2^2r_2^2)\ .
$$
By T-duality in $x_4$, we can also get another  D3-brane solution
with dilaton $e^{2\phi }=f(r_1,r_2)$.

Dimensional reduction of \mmmt\ along $x_4$ gives a D4 brane
solution of type IIA supergravity  with the metric
\eqn\fft{
ds^2_{10A}=H^{-1/2}\big[-dt^2 + dx_i^2  +dy^2\big] + H^{1/2}
\big[dr_1^2+r_1^2 (d\vp_1+b_1 dy) ^2+dr_2^2+r_2^2 (d\vp_2+b_2 dy)
^2+ dz^2\big]  \ ,
 }
 and the dilaton
 $e^{2\phi}= H^{-1/2}$.
For  the harmonic function being  $\vp_s$ and $z$ independent, 
e.g., 
$
H={L^2\over r_1^2+r_2^2}\ ,
$
this  solution should 
provide a dual description to SYM in 3+1 dimensions,
with  the number of unbroken supersymmetries ${\cal N}=0$ for
generic $b_s$, or ${\cal N}=2$ for $b_1=b_2$.

It would be interesting to study further the  low-energy
gauge theories corresponding to these solutions
(see in this connection \gan).

\listrefs
\vfill\eject
\end